\documentclass[journal,10pt]{IEEEtran}
\usepackage{amssymb}
\usepackage{amsmath}
\usepackage{cite}
\usepackage{url}
\usepackage{xcolor}
\usepackage{subfigure}
\usepackage{cite,graphicx,amsmath,amssymb}
\usepackage{subfigure}
\usepackage{fancyhdr}
\usepackage{mdwmath}
\usepackage{mdwtab}
\usepackage{caption}
\usepackage{amsthm}
\usepackage{setspace}
\usepackage{algorithm}
\usepackage{algorithmic}
\usepackage{makecell}
\usepackage{diagbox}
\usepackage{stfloats}
\usepackage{float}
\usepackage{picinpar}

\newtheorem{theorem}{Theorem}

\newtheorem{lemma}{Lemma}

\newtheorem{corollary}{Corollary}

\allowdisplaybreaks
\makeatletter
\def\ScaleIfNeeded{%
\ifdim\Gin@nat@width>\linewidth \linewidth \else \Gin@nat@width
\fi } \makeatother
\begin{document}
\title{Near-Field Beam Management for Extremely Large-Scale Array Communications}

\author{
Changsheng~You, Yunpu~Zhang, Chenyu Wu, Yong Zeng, Beixiong Zheng, \\  Li Chen, Linglong Dai, and A. Lee Swindlehurst

\thanks{C. You and Y. Zhang are with Southern University of Science and Technology; C. Wu is with Harbin Institute of Technology; Y. Zeng is with Southeast University and Purple Mountain Laboratories; B. Zheng is with South China University of Technology; L. Chen is with University of Science and Technology of China;  L. Dai is with Tsinghua University; and A. L. Swindlehurst is with University of California, Irvine.}
}

\maketitle

\begin{abstract}
Extremely large-scale arrays (XL-arrays) have emerged as a promising technology to achieve super-high spectral efficiency and spatial resolution in future wireless systems. The large aperture of XL-arrays means that spherical rather than planar wavefronts must be considered, and a paradigm shift from far-field to \emph{near-field} communications is necessary. Unlike existing works that have mainly considered far-field beam management, we study the new near-field beam management for XL-arrays. We first provide an overview of near-field communications 
and introduce various applications of XL-arrays in both outdoor and indoor scenarios.
Then,  three typical near-field beam management methods for XL-arrays are discussed: near-field beam training, beam tracking, and beam scheduling. We point out their main design issues and propose promising solutions to address them. Moreover, 
other important directions in near-field communications are also highlighted to motivate future research.
\end{abstract}

\section{Introduction}


Emerging applications such as extended reality, holographic video, and autonomous vehicles are driving the evolution of today's fifth-generation (5G) wireless systems to future six-generation (6G) versions, targeting more stringent performance requirements including  unprecedentedly high data-rates, super-high reliability, global coverage, ultra-dense connectivity, etc. These requirements, however, may not be fully achieved by existing 5G technologies, thus motivating the need for developing innovative technologies for 6G. For example, by exploiting the enormous spectral resources in TeraHertz (THz) bands, 6G systems will be able to integrate sensing, imaging, and communications functions. In addition, massive multiple-input-multiple-output (MIMO), which has been commercialized in 5G, is also evolving on its path to 6G. In particular,  \emph{extremely large-scale arrays} (XL-arrays) promise to significantly enhance spectral efficiency and spatial resolution  by increasing the number of antennas by another order-of-magnitude \cite{9617121,nf_mag,liu2023near,bjornson2021primer}.

\begin{figure}[t]
	\centering
	\includegraphics[width=0.47\textwidth]{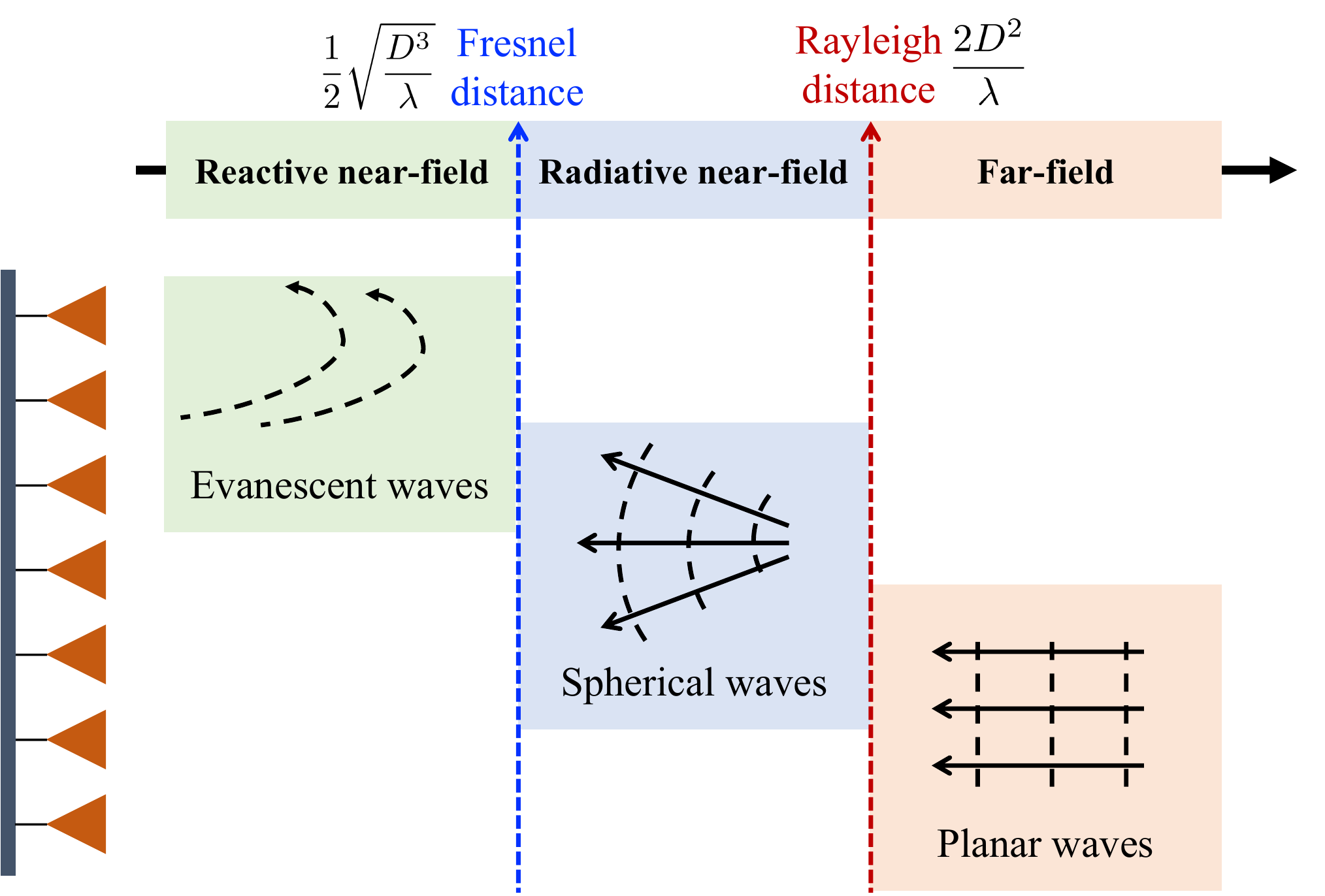}
	\caption{Illustration of different field regions, where $D$ and $\lambda$ denote the array aperture and carrier wavelength, respectively.} \label{Fig.NearReg}
	\vspace{+5pt}
\end{figure}

With the above technology trends of migrating to higher frequency and deploying more active/passive antennas, 6G will lead to a fundamental change in modeling electromagnetic (EM) propagation, shifting from the conventional far-field assumption to embrace \emph{near-field} communications. Taking the XL-array receiver as an example, the corresponding EM field can be roughly divided into three regions as shown in Fig.~\ref{Fig.NearReg}. 1) The reactive near-field region, where the transmitter is very close to the XL-array and within the {\em Fresnel distance} as specified in Fig.~\ref{Fig.NearReg}.  In this region, there are noticeable amplitude and and non-linear phase variations across the array aperture; 2) The radiative near-field (or Fresnel) region, where the transceiver distance is smaller than the  so-called \emph{Rayleigh distance}, the limit at which the signal propagation must be modeled by \emph{spherical wavefronts}. In this region, the amplitude variations across the antennas are negligible, while the phase variations are non-linear and can often be approximated as quadratic; 3) The far-field (or Fraunhofer) region with the transmitter located beyond the Rayleigh distance, in which the amplitude variations across the XL-array are negligible and the phase variations are linear. For instance, consider an XL-array with an  aperture of $D=0.4$ m and operating in a frequency of $f= 100$ GHz. The corresponding Fresnel and Rayleigh distances are $2.3$ m and $106.7$ m, respectively. The users of such an  XL-array system are very likely to be located in the radiative near-field region.

\begin{figure*}[t]
	\centering
	\includegraphics[width=0.93\textwidth]{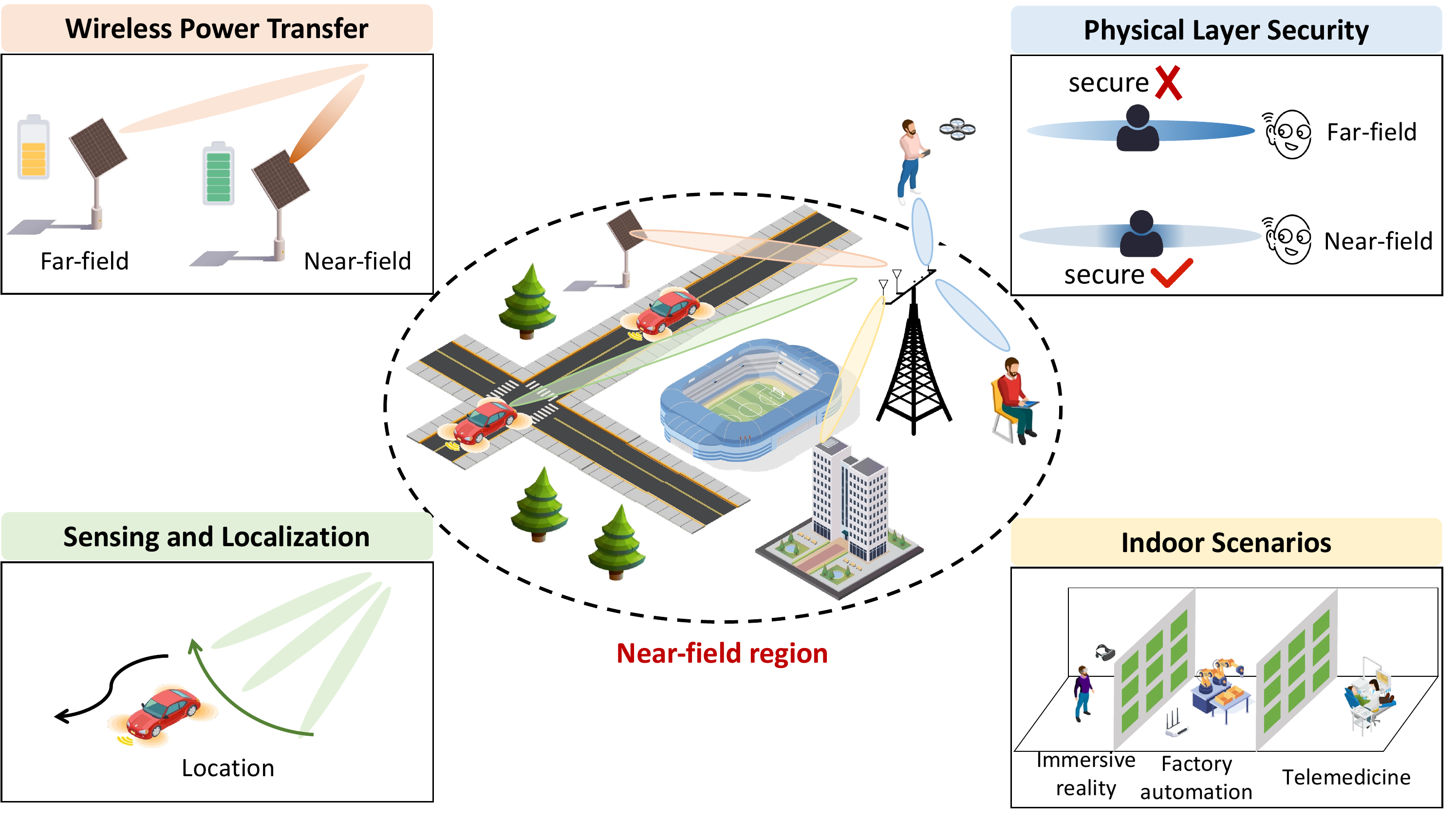}\\
	\caption{Typical application scenarios for near-field XL-array wireless systems.}
	\vspace{-12pt}
\end{figure*}

For  radiative near-field (hereafter referred to as simply near-field) communications, the  spherical wavefront characteristic provides several new opportunities. Specifically, unlike conventional far-field beamforming that steers beams with  planar wavefronts to specific \emph{angles}, in the near-field region, beams with spherical wavefronts can achieve  \emph{beam-focusing}, which concentrates the beam energy both in angle and range, thus providing a new degree of freedom (DoF) to control the spatial energy distribution.  By judiciously manipulating the beam-focusing regions, inter-user interference can be eliminated more effectively than in the far-field region \cite{9738442}. This provides an opportunity to serve multiple users located at similar angles but with different ranges, thereby significantly improving the spectral efficiency of multi-user communications. Furthermore, since the wave curvature over the XL-array is no longer negligible, the rank of even  line-of-sight (LoS) MIMO channels may be greater than one, which increases the opportunities for spatial multiplexing in support of multi-stream data transmissions.

In Fig. 2, we illustrate some typical applications of XL-arrays for improving the performance of wireless systems in both outdoor and indoor scenarios. For example, in  wireless power transfer (WPT) systems, XL-arrays can be used to reduce energy leakage by exploiting the beam-focusing property, hence greatly improving  power charging efficiency. Second,  XL-arrays can improve physical-layer security (PLS) by preventing eavesdroppers located in the same direction as the legitimate user from intercepting communications. With an XL-array at the base station (BS), signals can be intentionally focused on the legitimate user at a specific range, hence effectively reducing information leakage to an eavesdropper at a different range. In addition, XL-arrays can be used to enhance wireless sensing and localization performance by providing both angle and range information about the source of the received signals. Moreover, for indoor scenarios, XL-arrays can be  deployed to substantially increase achievable rates for accommodating data-intensive (e.g., immersive reality) and location-sensitive (e.g., telemedicine) applications.

Despite the appealing advantages and applications mentioned above, near-field XL-array wireless systems also face practical challenges. In particular, to exploit the beam-focusing gain of XL-arrays,  new and efficient beam management  methods need to be developed for the near-field case that are distinct from its far-field counterpart. This motivates the goal of the current work to provide an overview of near-field beam management for XL-arrays, including near-field beam training, beam tracking, and beam scheduling. In the following sections, we outline their main design issues and propose promising solutions to tackle them. Moreover, other important directions in near-field communications are also provided to inspire future work.

\section{Near-Field Beam Management}
\begin{figure*}[!t]
	\centering
	\subfigure[Near-field energy-spread effect.]{\label{fig:energy_spread}\includegraphics[height= 0.30\textwidth,width= 0.35 \textwidth]{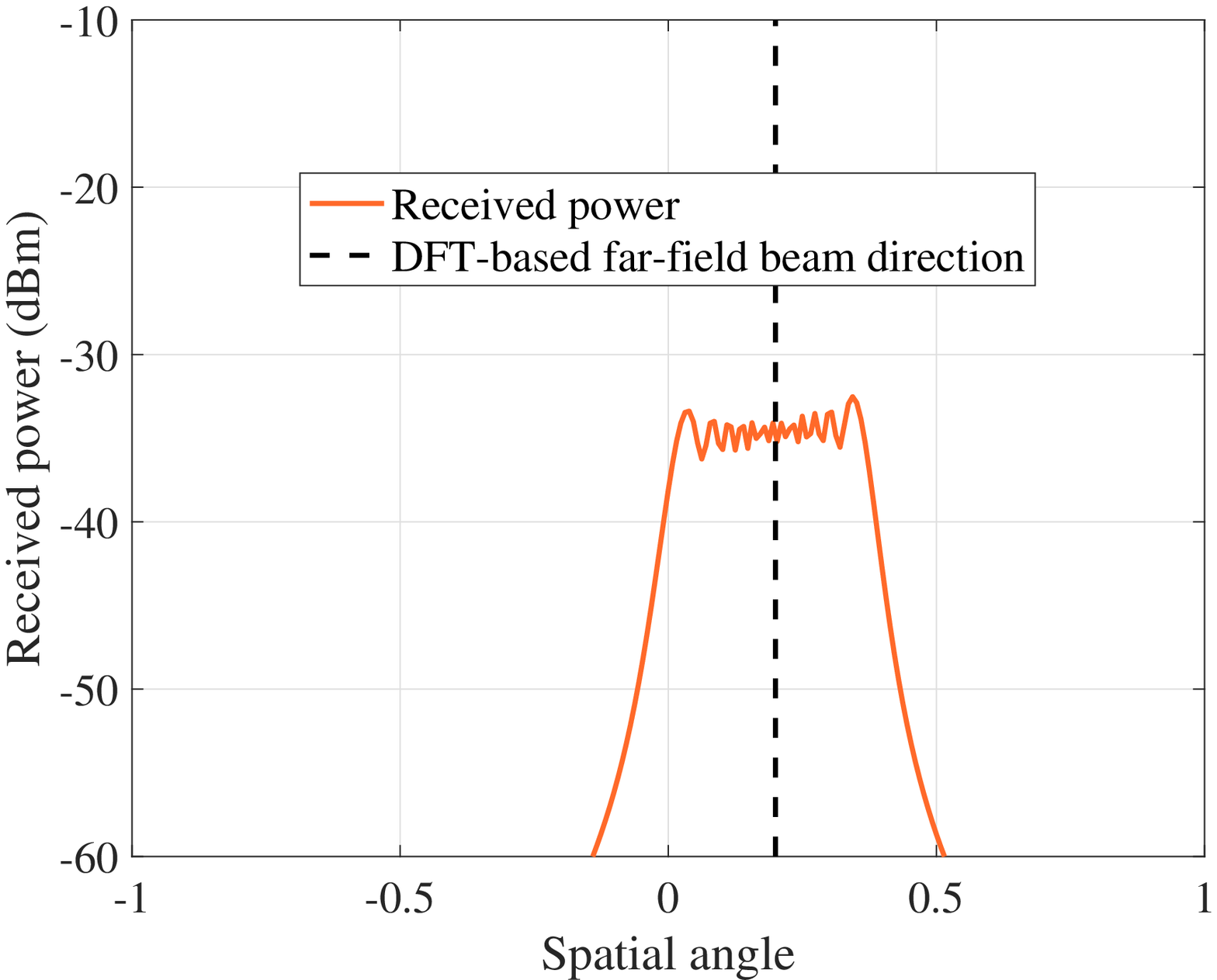}}
	\hspace{40pt}
	\subfigure[Near-field polar domain codebook.]{\label{fig:nf_beamtraining}\includegraphics[height= 0.30\textwidth, width=0.35\textwidth]{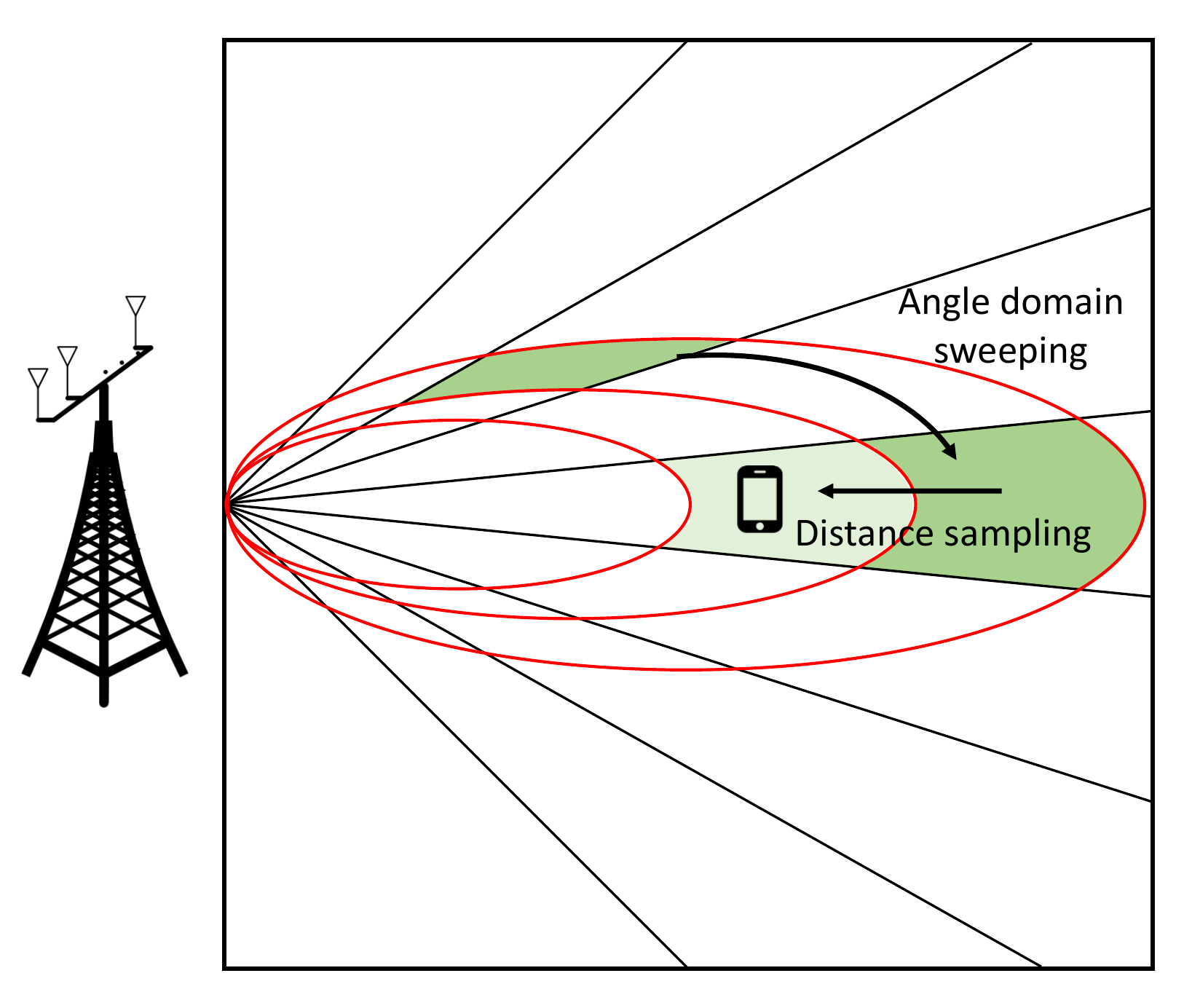}}
	\caption{Illustration of the near-field energy-spread effect and polar domain codebook.}
	\vspace{-12pt}
\end{figure*}
\subsection{Near-field Beam Training}

For XL-array communications in high-frequency bands, near-field beam training is of paramount importance, allowing the establishment of an initial high signal-to-noise (SNR) link between the BS and user before performing channel estimation and data transmission. Compared with conventional far-field beam training, near-field beam training faces several new challenges  due to its ability to discriminate users in the near field, as elaborated below. First, for uniform linear arrays (ULAs), unlike far-field beam training that only requires a one-dimensional beam search in the angular domain, near-field beam training requires a \emph{two-dimensional} (2D) beam search over both the angular and range domains. As such, directly applying the conventional discrete Fourier transform (DFT)-based far-field codebook for near-field beam training will inevitably result in degraded training  accuracy and rate performance. This is fundamentally due to the near-field \emph{energy-spread effect} illustrated in Fig.~\ref{fig:energy_spread}, in which the energy of a DFT codeword steered  towards a specific angle will be spread over a range of angles.  This effect renders it incapable of finding the true user angle based on the strongest received signal power. Therefore, new codebook designs tailored to near-field beam training need to be developed. For example, a 2D \emph{polar-domain} codebook was recently proposed in \cite{nf_exhaustive}, where each beam codeword points to a specific \emph{location} with target angle and distance. Moreover, in terms of the angle and distance sampling, it was shown that the column coherence of adjacent codewords can be minimized by uniformly sampling the angular domain, while  the range domain requires \emph{non-uniformly} sampling with the sampling density decreasing with the distance, as illustrated in Fig.~\ref{fig:nf_beamtraining}. This is because  phase variations due to range are  more pronounced when the distance becomes shorter, while the spherical wavefront eventually becomes planar when the distance is sufficiently large.

Given a polar-domain codebook, a straightforward beam training method is to apply a 2D  exhaustive  search over all possible beam codewords. This, however, will incur prohibitively high beam training overhead proportional to the product of the number of sampled angles and distances, denoted by $N$ and $S$, respectively,  thus leaving less time for data transmission. To reduce this overhead, a simple yet efficient \emph{two-phase} near-field beam training method was proposed in \cite{two_phase}. This method exploits a fact 
that when the DFT codebook is used, the true user angle approximately lies in the middle of a dominant angular region. As such, the best near-field beam codeword can be efficiently identified by first estimating the user angles using the DFT codebook and then resolving the user ranges using the polar-domain codebook. 
Although this method greatly reduces the training overhead to be proportional to $N+S$, $N$ may still be very large for XL-arrays (e.g., $N=1024$). 

The above  motivates the design of hierarchical near-field beam training methods to further reduce the overhead. To this end, new  near-field hierarchical  codebooks need to be devised that possess the following two properties: 1)
an elementary sub-codebook that covers the entire polar domain;  and 2) step-wise sub-codebooks that guarantee the beam pattern of upper-layer polar-domain codewords being approximately covered by multiple codewords at a lower layer. Based on this, the user angle and range
 can be simultaneously updated with a finer resolution, hence achieving a smaller training overhead that scales only as $O( \log{N}+\log{S}$) \cite{9941256}. 
Another possible approach to address this issue is to employ a portion of the XL-array to estimate a coarse user angle using a conventional far-field hierarchical  codebook, and then progressively resolve fine-grained user angle and range in the polar domain \cite{wu2023two}. However, this method may suffer a performance loss in the low-SNR regime since it only uses part of the XL-array for the coarse angle search. Furthermore, for wideband XL-array systems, the near-field \emph{beam-split} effect poses new challenges for beam training, where the beams generated at different frequencies are distorted to different locations. To tackle this challenge, an effective wideband rainbow beam training scheme was proposed in \cite{9957130}, which utilizes true-time-delay (TTD) devices to flexibly control the near-field beam-split effect for fast beam training. Specifically, in each training interval, multiple sub-beams are simultaneously generated towards different locations with the same range but different angles, while the target ranges are dynamically modified over time to find the best polar-domain codeword, thereby achieving a small overhead proportional to only $S$.

\begin{figure*}[!t]
	\centering
	\subfigure[Near-field beam tracking with a single XL-array.]{\label{fig:beam_tracking1}\includegraphics[height= 0.28\textwidth]{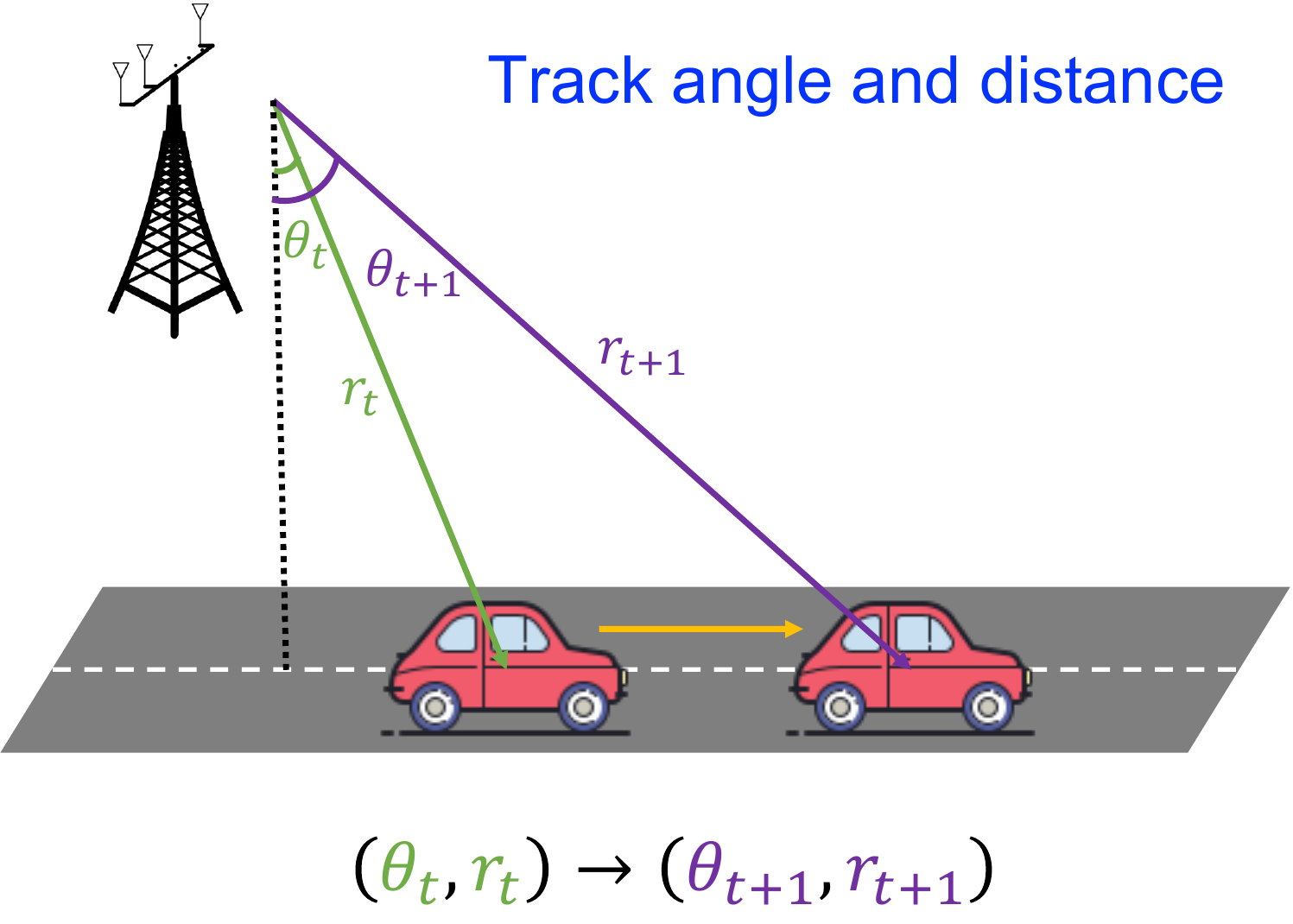}}
	\hspace{35pt}
	\subfigure[Collaborative near-field beam tracking with two XL-arrays.]{\label{fig:beam_tracking2}\includegraphics[height= 0.28\textwidth]{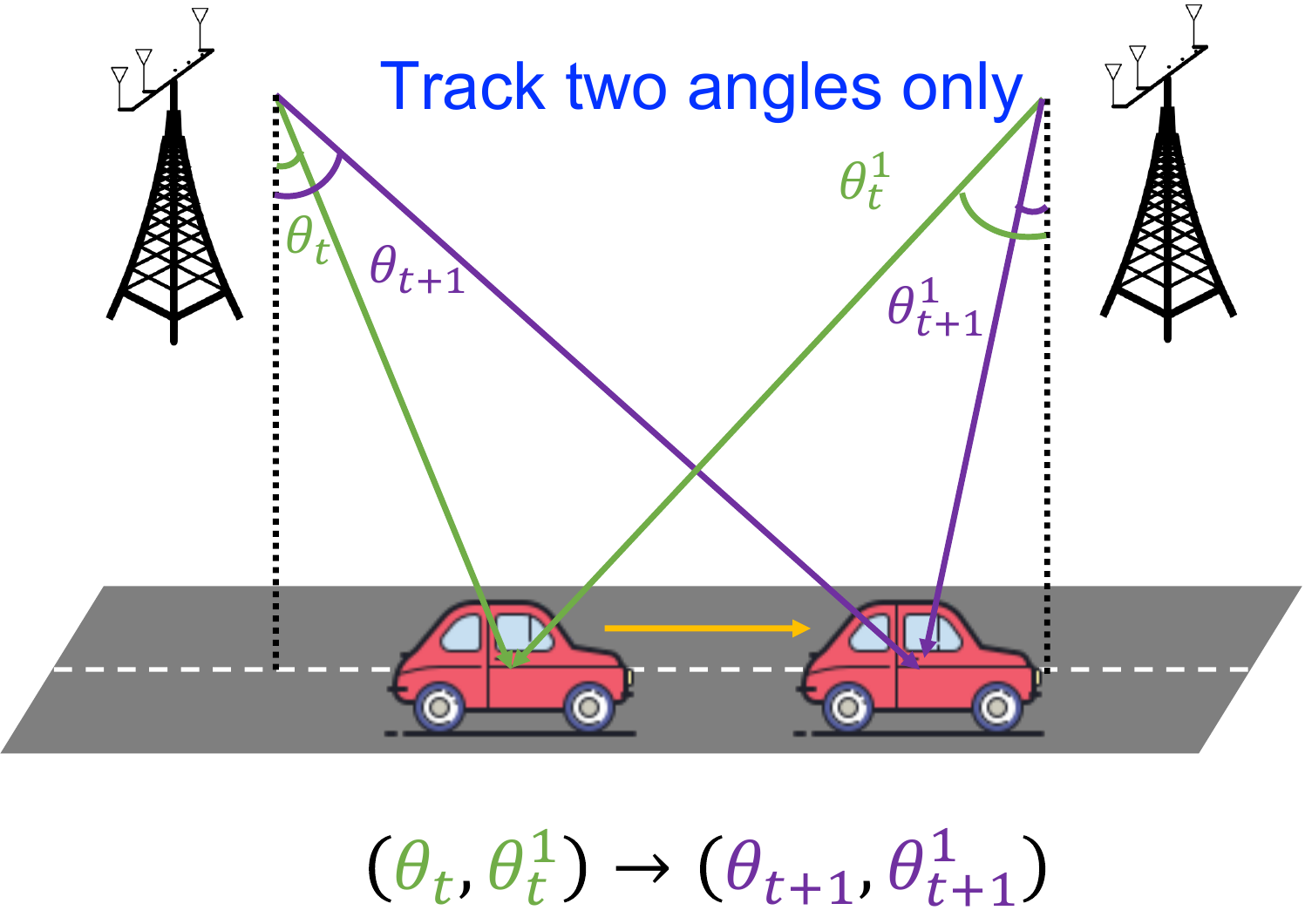}}
	\caption{Different near-field beam tracking methods.}
	\vspace{-12pt}
\end{figure*}

Several challenges in near-field beam training remain  unsolved.
For example, most  existing works  considered a single LoS path with a spherical wavefront, while the more general channel model with multiple additional non-LoS (NLoS) paths in the near-field region has not been studied, a situation that is particularly important when there are multiple scatterers between the XL-array and user. A possible solution is to first identify the angles of all paths by examining the dominant angular regions using the DFT codebook and then use the polar-domain codebook to determine the scatterers' distances one by one. However, this method may not work well when the dominant angular regions of different paths significantly overlap. In addition, it would also be of interest to investigate  how to extend existing near-field beam training methods for ULAs to other  array structures such as a uniform planar array (UPA) or uniform circular array (UCA). Furthermore, how to design a \emph{unified} beam training method suitable for both near-field and far-field scenarios is a crucial problem, since users may be distributed in both  the near- and  far-field regions in certain scenarios. This requires further in-depth research on the effective fusion of near- and far-field beam training methods. Another possible approach can be first quickly identifying whether the user is located in the near- or far-field, and then applying the corresponding beam training methods. Last but not least, the visibility region (VR) of XL-arrays may also affect the near-field beam training performance in practice, where different users may observe different regions of the XL-array \cite{8949454}. How to perform efficient multi-user beam training and possibly reduce the training overhead in the presence of non-uniform visibility regions is an  interesting problem worthy of future investigation.

\vspace{-5pt}

\subsection{Near-field Beam Tracking}

For high-mobility near-field users, in addition to beam training, efficient beam tracking methods also need to be devised to maintain their high-quality links with the XL-array BS over time. Otherwise, slight beam misalignment in the near field could incur severe performance loss with the energy-focusing effect taken into account. This issue is exacerbated when the number of antennas increases due to the reduced beamwidth. This thus requires the design of accurate, fast, and robust near-field beam tracking methods.

Near-field beam tracking is more challenging than conventional far-field beam tracking, as elaborated below. First, unlike far-field beam tracking that only tracks user angles over time, near-field beam tracking needs to dynamically adjust its beam towards both the user angle and range based on the user trajectory. Also, as mentioned above, XL-arrays require a large-sized polar-domain codebook, which  generally incurs high overhead. Additionally,  near-field beam tracking performance is more sensitive to user motion, as such motion tends to cause faster phase changes across the XL-array aperture than in the far-field case.
Consequently, directly applying existing far-field beam tracking methods will cause considerable performance degradation, thus calling for the design of new beam tracking methods dedicated to near-field communications.


First, consider near-field beam tracking with a single XL-array as shown in Fig.~\ref{fig:beam_tracking1}.  When the user mobility model is known \emph{a priori}, methods based on Bayesian statistics can be utilized to track the optimal beam. For example, if the channel dynamics can be modeled as  a non-linear Gaussian process, the extended Kalman filter  (EKF)  can be applied to predict the best beam based on estimated and predicted user positions and velocities. Moreover, if a kinematic model is given, the unscented Kalman filter (UKF)  can be utilized to improve beam tracking accuracy. However, the performance of these methods critically depends on the accuracy of the user mobility model, and an inaccurate or outdated model can result in poor performance. 
In addition, near-field beam tracking also becomes more difficult if the user follows an irregular trajectory over time. On the other hand, useful side information can also be exploited to predict the best beam angle and/or distance for near-field beam tracking, hence effectively reducing the beam search space. For example, one can first construct offline a channel knowledge map that includes reference locations of some anchor/sensor nodes and their best beam codewords. 
 Then, in the online beam tracking stage, the XL-array can first estimate the user angle (or range) based on these nearby anchor/sensor nodes, and then find the best polar-domain beam among a small number of shortlisted candidate codewords. The key challenge is how to quickly identify the best beam based on available side information such as approximate user angle and/or distance. In general, the number of shortlisted beams is determined by several factors such as user speed, angle, position, and the tracking accuracy requirement. The higher the required beam training accuracy, the greater the number of candidate beams that need to be tested. In addition, it is also important to properly allocate limited beam resources to near-field beam tracking in the  angular and range domains.
Next, consider \emph{collaborative} near-field beam tracking with multiple XL-arrays, as depicted in Fig. ~\ref{fig:beam_tracking2}. In this case,  each XL-array BS can track the user angle over time using e.g., EKF based methods; then, the user location can be effectively estimated from the XL-array positions and the relative user angles between them, as shown in the figure. This method avoids time-consuming distance tracking, as the intersections of the lines connecting each XL-array and the user can be used to identify user location, thereby circumventing the training overhead for range estimation.

\begin{figure*}[!t]
	\centering
	\subfigure[Near-field beam scheduling.]{\label{fig:beam_scheduling1}\includegraphics[height= 0.24\textwidth]{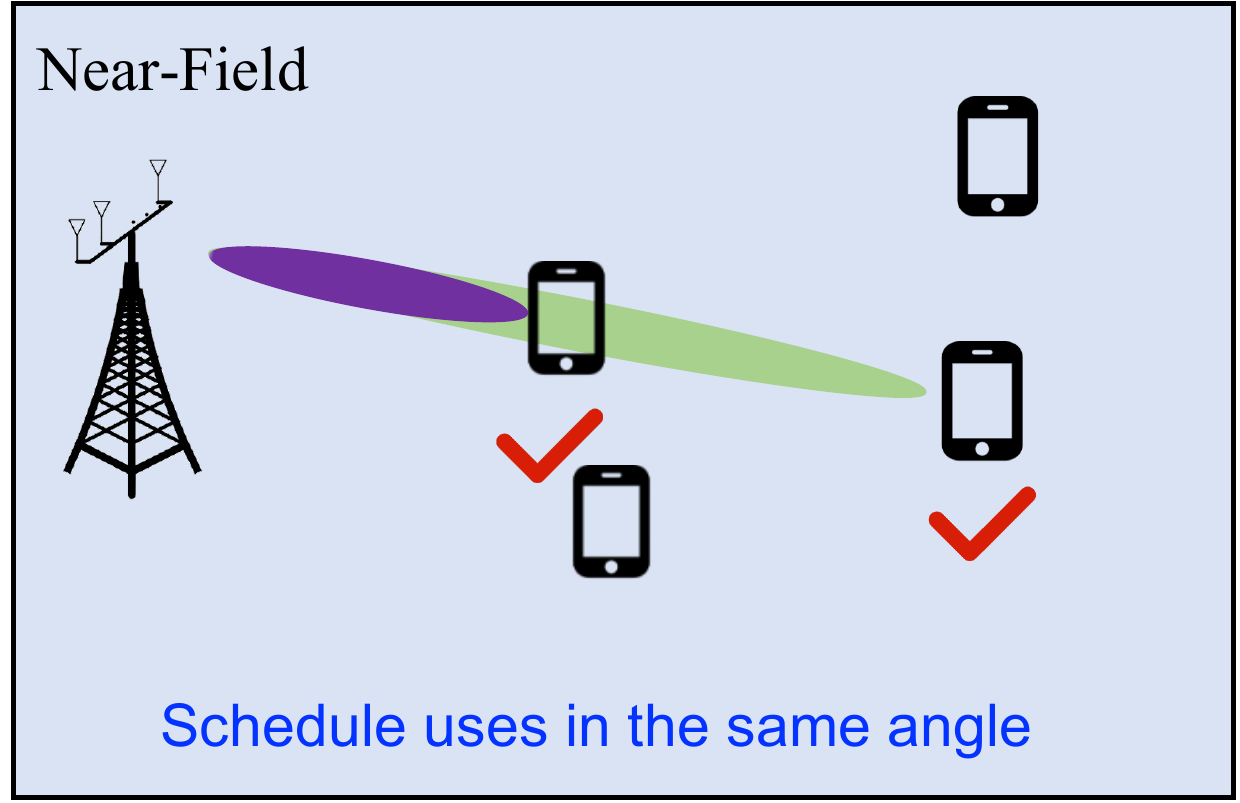}}
	\hspace{55pt}
	\subfigure[Mixed-field beam scheduling.]{\label{fig:beam_scheduling2}\includegraphics[height= 0.24\textwidth]{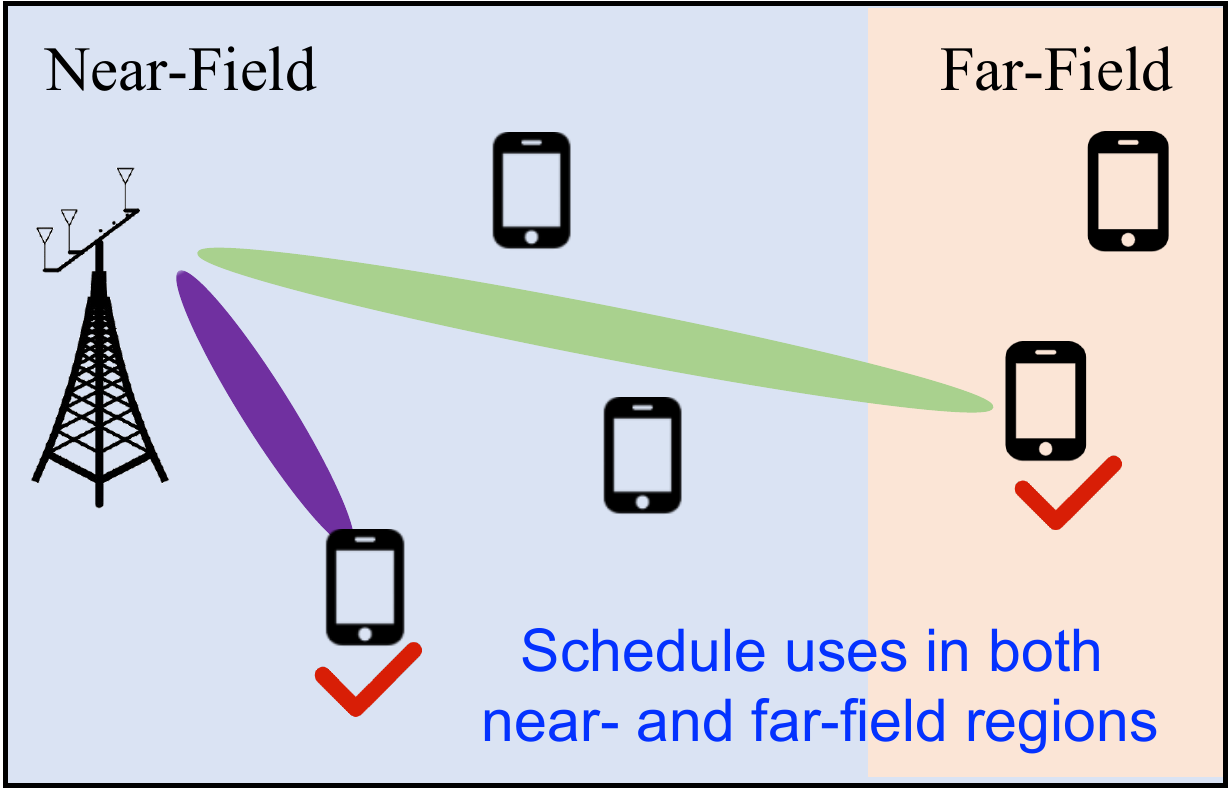}}
	\caption{Beam scheduling in different regions.}
	\vspace{-12pt}
\end{figure*}

In practice, the performance of near-field beam tracking is  affected by various factors such as receiver noise, measurement noise, estimation error, etc. Moreover, for high-mobility users or rapidly-changing trajectories, it is likely that the signal  deviates temporarily from the beam mainlobe and hence causes intermittent tracking failures. To address these issues, robust beam tracking methods need to be developed to improve reliability. A promising approach is to use  the particle-filter based beam tracking methods to sequentially estimate the user's physical parameters (e.g., angle, range, channel gain). Then the misalignment probability can be estimated from, for example, deviations in the user angle and distance, based on which the beamwidth of the XL-array can be adaptively adjusted in real time. 
Additionally, auxiliary beam pairs can be provided to increase the probability of successful beam alignment.

\vspace{-5pt}

\subsection{Near/Mixed-Field Beam Scheduling}
For conventional far-field beam scheduling, the transmit beam is  usually tuned to the user angle to maximize  the received signal power. However,  for near-field beam scheduling, both the user angle and \emph{range} information should be considered in the design of joint multi-user beam scheduling and power allocation.
In particular, the beam-focusing property  introduces  range-domain resolution in near-field communications, which provides additional DoF in the beam scheduling design, as elaborated below.
 

For  far-field communications, spatial division multiple access (SDMA) or beam division multiple access (BDMA) can be employed to simultaneously serve multiple users at different angles with small/no inter-user interference, thanks to the asymptotic orthogonality of the far-field beam vectors. However, SDMA/BDMA may not work well when users are located at the same angle due to the strong inter-user interference. Interestingly, this problem can be alleviated in near-field beam scheduling when the number of XL-array antennas becomes sufficiently large. 
Specifically, as illustrated in Fig.~\ref{fig:beam_scheduling1}, the near-field beam-focusing effect allows multiple users to be served simultaneously at the same angles, achieving
 enhanced accessibility without causing severe inter-user interference, a feature that is referred to as \emph{location division multiple access} (LDMA). For LDMA, a key challenge is how to design a range-and-angle aware beam scheduling scheme. For example, consider the near-field beam scheduling problem for maximizing the number of accessible users subject to a minimum rate constraint for each individual user.
 In this case, it is desirable to simultaneously schedule users with small inter-user inference in the angular and/or range domains. Furthermore, when the XL-array is physically large, the VRs of different users need to be considered in beam scheduling. For instance, when an XL-array is divided into multiple sub-arrays, users in non-overlapping VRs can be scheduled at the same time using different sub-arrays, while users with partially-overlapped VRs can be scheduled separately to reduce inter-user interference.

It is also of interest to consider \emph{mixed} (or hybrid) near-field and far-field communication scenarios for XL-array systems, in which
some users are located in the near field and the others reside in the far field. This gives rise to several new design issues, such as joint beamforming, scheduling, and power allocation. First, consider analog beamforming when an XL-array steers multiple  beams towards individual near-field and far-field users.  The near-and-far inter-user interference is closely related to the correlation between near-field and far-field steering vectors. In particular, an interesting result from  \cite{zhang2023mixed} is that when the angle of a far-field user is similar to  one in the near field,  the near-and-far channel correlation is relatively  high even if the angles are not identical. Since the BS usually allocates more transmit power to  far-field users to ensure user fairness, near-field users generally suffer greater interference from the far-field beams, while far-field users are less affected by the near-field beams due to the smaller signal power and more severe path-loss. 
Therefore, for  mixed-field multi-user communication systems, efficient beam scheduling and power allocation approaches as well as non-orthogonal multiple access (NOMA) techniques need to be designed to achieve a balance between maximizing desired signal power and minimizing mixed-field interference (see Fig.~\ref{fig:beam_scheduling2}). 

Near-and-far field correlation can also be exploited to improve the performance of mixed-field simultaneous wireless information and power transfer (SWIPT) systems, where energy harvesting (EH)  and information decoding (ID) receivers are located in the near- and far-field of the XL-array, respectively. Specifically, it was shown in \cite{zhang2023joint} that 
by exploiting the energy-spread effect, near-field EH receivers can be efficiently and opportunistically charged by the DFT-based beams directed towards far-field users when they are located at similar angles. Moreover, the near-field beam-focusing effect can be leveraged to enhance near-field EH efficiency, while at the same time minimizing the interference to far-field ID receivers. In particular, the authors of \cite{zhang2023joint} showed that when there are multiple EH receivers and one ID receiver, the optimal design in most cases is to schedule together the far-field ID receiver and a near-field EH receiver for maximizing the harvested power under a rate constraint, which is in sharp contrast to the conventional far-field SWIPT case, for which all power should be allocated to the ID receiver.

\section{Numerical Results}
Numerical results are presented in this section to demonstrate the effectiveness of near-field beam training designs. We consider an XL-array communication system in which the XL-array at the BS has $N=512$ antennas and the carrier frequency is $f=100$ GHz. We define the reference SNR as ${\rm SNR}=\frac{P\beta N}{r^2 \sigma^2}$  with transmit power $P=30$ dBm, noise power $\sigma^2=-80$ dBm, and reference path-loss $\beta=-72$ dB. To characterize the beam 
determination accuracy, the beam-training
success rate is defined as the probability of finding the best beam codeword in the polar-domain codebook.  For performance comparison, we consider the following beam training schemes: 1) \emph{perfect channel state information (CSI)  beamforming,} which perfectly aligns the transmit beamformer with the user channel and thus serves as a performance upper bound; 2) \emph{exhaustive-search near-field beam training,} which applies a 2D exhaustive search to find the best polar-domain codeword \cite{nf_exhaustive}; 3) \emph{two-phase near-field beam training}, which sequentially estimates  the user angle and range \cite{two_phase};  4) \emph{two-stage hierarchical near-field beam training,} which first searches for a coarse user angle with a conventional far-field hierarchical codebook and then progressively refines the user angle and range  using a near-field hierarchical  codebook \cite{wu2023two}; and 5) \emph{far-field hierarchical beam training,} which directly uses the far-field hierarchical codebook for beam training.

\begin{figure}[t]
\centering
\subfigure[Beam training success rate and overhead with different schemes.]{\label{Fig:SR}
\includegraphics[width=7.6cm]{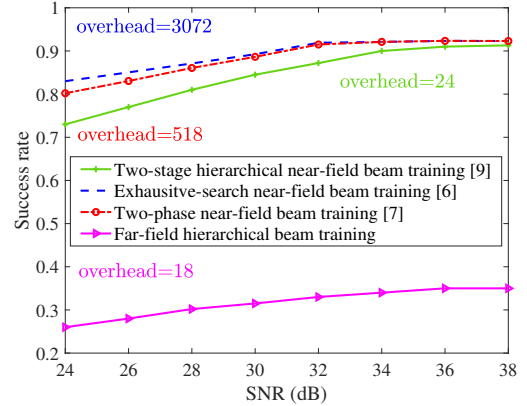}}
\hspace{5pt}
\subfigure[Achievable rate with different beam training schemes.]{\label{Fig:AR}
\includegraphics[width=7.6cm]{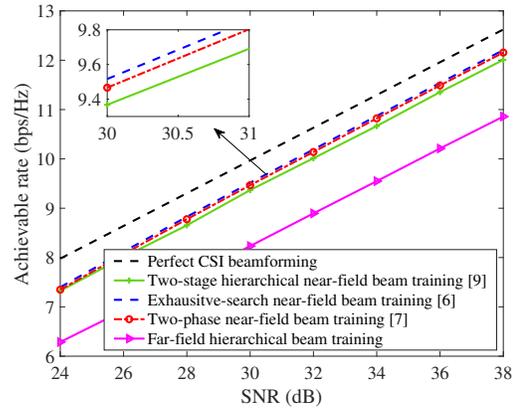}}
\caption{Performance comparison of different beam training schemes.}
\end{figure}

In Fig.~\ref{Fig:SR}, we show  the success rates and training overhead of  different beam training schemes versus the reference
SNR. It is observed that the success rates for all approaches monotonically increase with the SNR. Second, the two-phase near-field beam training method achieves a success rate very close to that of the near-field exhaustive search, albeit with much lower training overhead ($518$ versus $3072$ beam training symbols). Moreover, the training overhead is  further significantly reduced in the two-stage hierarchical beam training approach (i.e., $24$ versus $518$),  without suffering much loss in terms of success rate. Although the far-field hierarchical scheme has the lowest training overhead, its success rate drops significantly  due to channel model mismatch. 
In Fig.~\ref{Fig:AR}, we plot  the achievable rates  of the different approaches versus the SNR. Again, it is observed that the two-phase \cite{two_phase} and two-stage hierarchical \cite{wu2023two} near-field beam training schemes achieve rate performance comparable to the exhaustive-search approach, and both significantly outperform  far-field beam training.

\vspace{-2pt}
\section{Conclusions and Future Directions}

In this article, we have provided a comprehensive overview of near-field beam management for XL-arrays. Specifically, we first presented unique characteristics of near-field communications and introduced practical applications of XL-arrays. Then, we gave an overview of near-field beam management, including beam training, beam tracking, and beam scheduling. We discussed their main design issues as well as promising solutions. Some other directions for future research in near-field communications are outlined as follows.

\textbf{Channel modeling:} 
Channel modeling lays the fundamentals for near-field XL-array communications. In existing works,   near-field LoS channel models have been widely assumed. However, more practical and general channel models for near-field communications need to be studied. For example, it is crucial  to study  more complex multipath channels for near-field communications that consider multipaths due to  
 ambient environmental scatterers in the far-field and/or near-field of the XL-array  \cite{ZengySpatial}. The environmental scatterers also affect the VRs for different users, and this effect needs to be properly modeled. Furthermore, in addition to deterministic channel models, stochastic near-field channel models  need to be established to account for near-field spatial correlation and non-stationarity. 

\textbf{Machine learning:} Beamforming optimization and beam management for near-field communications are more challenging than in the far-field case  due to the spherical-wave propagation with non-linear phase variations. 
To reduce the training overhead and beamforming design complexity, deep learning/reinforcement learning methods can be invoked  to learn the non-linearity in the near-field channels as well as the mapping between  CSI and optimal beamforming for multi-user communications \cite{9903646}. However, the robustness and design of suitable machine learning frameworks are important open problems worthy of  future research.
  
\textbf{Hardware-efficient transceivers:} Since near-field XL-array communications usually operate in high-frequency bands, high power consumption and hardware complexity become core concerns. A promising solution is exploiting hybrid beamforming to reduce the hardware and energy cost. In addition, sub-connected architectures, dynamic subarray architectures, and lens antenna arrays can also be employed and properly designed to balance the tradeoff between complexity and performance. Next, phase-shifter based analog components cannot handle the beam split effect in  wideband near-field communications.
An effective solution is to employ additional TTD circuits between the radio-frequency chains and phase-shifters  to generate frequency-dependent phase shifts and hence focus the beam over the entire bandwidth. This direction is still in its early stage and deserves further investigation. 

\textbf{Integrated designs with other promising technologies:} 
\begin{itemize}
	\item \textit{Intelligent reflecting surface (IRS):} For XL-IRSs operating in high-frequency bands, near-field channel modeling needs to be considered and will lead  to new design principles for IRSs. First, optimal XL-IRS placement needs to be reconsidered, depending on whether the BS or users or both are located in the near- or far-field of the XL-IRS.  Second, if both the BS and users are in the near-field of the XL-IRS,  IRS beam training becomes more complicated because the phase of the received signals is accumulated by the propagation delays through both the BS-RIS and RIS-user links. As such, future studies need  to consider  this \textit{double-sided} near-field effect to achieve superior beam training performance. 		
	\item \textit{Integrated sensing and communications (ISAC):} ISAC has emerged as a key technology for 6G, since it not only unifies these two functions in one system, but also realizes flexible performance tradeoffs and mutual performance gains. For near-field ISAC, opportunities and challenges coexist. On the one hand, the near-field spherical-wave propagation model introduces a new DoF in the range domain, which offers the potential to improve wireless/radar sensing performance via joint angle and range sensing. On the other hand, near-field ISAC requires new waveform designs to balance the tradeoff between  near-field sensing and communications. 	
Moreover, the fundamental limits of near-field communication and sensing performance are still  unknown and require future investigation.
\end{itemize}

\bibliographystyle{IEEEtran}
\bibliography{ref}
%
%
%
%
%
%
%
%
%
%
%
%
%
%
\end{document}